\documentclass[%
 reprint,
 amsmath,amssymb,
 aps,
]{revtex4-2}

\usepackage{graphicx}
\usepackage{dcolumn}
\usepackage{bm}
\usepackage{color}

\begin{document}

\preprint{APS/123-QED}

\title{Hidden Signals of New Physics within the Yukawa Couplings of the Higgs boson}

\author{M. A. Arroyo-Ure\~na}
 \email{marcofis@yahoo.com.mx}

\affiliation{%
  Departamento  de  F\'isica,  FES-Cuautitl\'an,\\
Universidad  Nacional  Aut\'onoma  de  M\'exico,\\
C.P.  54770,  Estado  de  M\'exico,  M\'exico. 
}%
\author{J. Lorenzo Diaz-Cruz}%
 \email{jdiaz@fcfm.buap.mx}
 \affiliation{Mesoamerican Center for Theoretical Physics, \\ 
 Universidad Autónoma de Chiapas (M\'exico) \\
 and Facultad de Ciencias F\'isico-Matem\'aticas\\
Benem\'erita Universidad Aut\'onoma de Puebla,  Puebla, Pue., M\'exico.
}%


\date{\today}

\begin{abstract}
The Large Hadron Collider (LHC) has measured  the Higgs boson couplings with the heavier particles of the Standard Model (SM), and they seem to lay on a single line as function of the particle mass, as  predicted in the SM. However a complete test of this property must involve the lighter generations, and the coming measurements at LHC,  or future colliders, may reveal hidden patterns associated with physics beyond the SM. In renormalizable multi-Higgs models, the Higgs-fermion couplings could still be linear, but they could lay on multiple lines.
Then, the angle subtended by  these lines, with respect to the SM line,
 can be used to characterize different models,  as it is shown here for the Two-Higgs doublet Model (2HDM). Models where fermion masses arise from higher-dimensional operators, may also result in large deviations from the SM values for the Higgs couplings, which would show an irregular pattern as a function of the fermion masses. In the case of neutrino masses, when they arise from the see-saw mechanism, one finds that the Higgs Yukawa couplings will show a non-linear mass relation.




\begin{description}
\item[PACS: 12.10.Dm, 12.60.-i , 14.80.Cp]
\end{description}
\end{abstract}

\maketitle





\section{Introduction \label{Introdcution}}
The physics of the Higgs boson post-discovery \cite{lhc1,lhc2} has entered a new level of precision, 
with new decay modes and properties expected to be studied at the LHC \cite{lhc3}. 
Within the SM only  one Higgs doublet gives masses  to gauge bosons 
and all type of fermions, thus their couplings  are proportional to the  particle mass, 
and  when plotted as function of the masses, they lay on a single line. 
  In particular,  the Yukawa couplings of the Higgs boson with the third generation fermions ($t, b, \tau$), seem to be
 in agreement with this SM relation. Furthermore, the LHC has started to probe the 
Higgs couplings for the lighter generations \cite{lhc3}, which will open the possibility  to test this SM 
 prediction with better precision. Different patterns or relations associated with models of new physics, could be hidden 
in the data, and this may show up only after using the appropriate Higgs observables.
Then, one could compare the predictions for these observables coming from different 
extensions of the SM Higgs sector \cite{Branco:2011iw}.

A variety of signal strengths $\mu_X$  for different channels have been measured at LHC, these include the Higgs boson production through gluon fusion, in association with a gauge boson ($W,Z$) or with top quark pairs, and with the Higgs boson decays into $X\bar{X}$, with $X=b,\,\tau,\,\mu,\,W,\,Z,\,\gamma$. The SM Higgs boson couplings for third generation fermions and gauge bosons, have been extracted from these channels.
Moreover, it is expected that LHC will be able to probe the Higgs couplings with second generation fermions,   
with different levels of precision.  
 The LHC bounds on the signal strengths for the Higgs decay into muons is already reaching $O(1)$ level (more next),
 while for charm and  strange quarks the corresponding  bounds from current LHC runs are 
 $\mu_{cc} =110$ and  $\mu_{ss}=200$, respectively; while theory estimates at future LHC-HL are $\mu_{cc} =16$,   $\mu_{ss}=100$
 \cite{Delaunay:2013pja,Gray:2019mdt}.

In this letter we look at the dependence of the Higgs couplings on the fermion masses in models with extended Higgs sector.
We shall consider first renormalizable multi-Higgs models, which also display a linear relations for the Higgs couplings \cite{Cruz:2019vuo}. 
 When more than one Higgs multiplet participates in the fermion mass generation, the single Higgs Coupling Line (\textbf{HCL})  
 obtained from LHC Higgs boson data, would only be apparent, and when looked more closely it would  reveal some sub-structure. For instance, within the two-Higgs doublet model types,  the fermion couplings could lay on a single line (2HDM-I) or two lines (2HDM-II).
 There are also models where each fermion type gets its mass from a private Higgs \cite{Porto:2007ed,Diaz-Cruz:2014pla};
 for the realization of this idea, one needs at least 3 Higgs doublets, (4 doublets in the SUSY case) \cite{Aranda:2012bv,Diaz-Cruz:2019emo}; 
 here the Higgs boson couplings would lay on three different lines.
In all these models, the angle subtended  by those \textbf{HCL}  and the SM one, can be calculated in terms of the fermion masses and Higgs couplings, then one can study their behavior in the alignment or decoupling limits. 

We could also consider models where fermion masses arise from higher-dimensional operators, which have been considered as an attempt to explain the hierarchies in the flavor parameters (masses and mixing angles) with an scale of $\mathcal{O}$(TeV) \cite{Babu:1999me}. 
In this case  the dependence of the Higgs boson couplings on fermion masses would still be linear, but they would not appear laying on defined lines,  rather they would look as dispersed points. For neutrinos the situation is less clear,  it  depends on the mechanism of mass generation (see-saw of type I,II, III, etc), or weather it is of Dirac or Majorana type \cite{King:2019gif, Valle:2020wdf}. As it will be discussed at the end of our paper, when the neutrino masses are of the Dirac-type, the Higgs couplings would be linear in the neutrino masses, but when they arise from the see-saw mechanism, the corresponding Higgs boson couplings would obey a non-linear mass relation.  




\section{Higgs Couplings in Renormalizable Multi-Higgs Models\label{Sec2}}
Let us study an extension of the SM with $n$ Higgs doublets $\Phi_a = (\Phi^+_a, \Phi^0_a)^T$, with $a=1,\,2,...,n$, and write the Yukawa Lagrangian for the fermion types $f=u,\,d,\,\ell$ as follows:
\begin{equation}\label{LagrangianMultiHiggs}
{\cal{L}} = \sum^n_{a=1} {\bar{F}}_{Li} \Phi_a \lambda^a_{ij} f_{Rj} + h.c.
\end{equation}
It involves the left-handed fermion doublet $F_{Li}= (f'_{i,L}, f_{i,L})^T$ (with $f'_{i,L}, f_{i,L}$ denoting the fermions with isospin $T_3=+1/2,-1/2$, respectively) and right-handed fermion singlet $f_{Rj}$, with $i,j=1,2,3$ labeling the fermion generations. Here $\lambda^a_{ij}$ denotes the matrix of Yukawa couplings. Further, let us consider  the class of multi-Higgs models that respect the Glashow-Weinberg theorem \cite{Glashow:1976nt}, where each fermion type $f$ couples only with one Higgs doublet. For the up-type quarks $f=u$, we shall assume that it only couples to $\Phi_1$. After spontaneous symmetry breaking (SSB), i.e. $<\Phi^0_1>=\upsilon_1/\sqrt{2}$, the interaction of the neutral component of $\Phi_1$
coming from eq. \eqref{LagrangianMultiHiggs} is written as follows:
\begin{equation}
{\cal{L}} =  \lambda_{u,ij}  \bar{u}_{Li} u_{Rj} ( \frac{v_1+\phi^0_1}{\sqrt{2}}) + \sum^N_{a=2} {\bar{F}}_L \Phi_a \lambda^a f_R + h.c.
\end{equation}
The fermion mass is  then given by $m_u= \lambda_u v_1 / \sqrt{2}$, while the coupling of the fermion with $\phi^0_1$ is $g_{\phi_1 uu} = m_u/ v_1$, i.e. it is a linear relation as a function of the mass. As $\phi^0_1$ is not a mass eigenstate in general, one need to rotate to the Higgs mass basis, which will introduce elements of the rotation matrix into the corresponding Yukawa Couplings. Since we are assuming that $\Phi_1$ only couples to the up-type quarks, 
then the remaining Higgs doublets could at most couple to $d$-type quarks or leptons $\ell$. Specifying which Higgs doublet couples to a given
fermion type, defines the particular multi-Higgs model.

Thus, for these multi-Higgs models with natural flavor conservation, the Higgs couplings as function of the masses, could lay on one or more \textbf{HCL}. For instance, in the 2HDM-I, all charged fermions obtain their masses from one of the Higgs doublets, say $\Phi_1$, 
and all the Yukawa couplings will lay on a single line, as in the SM. But in general, this line will not coincide with the SM one. 
In the 2HDM-II, one Higgs doublet ($\Phi_2$ within our conventions) gives masses to $d$-type quarks and leptons, and the other ($\Phi_1$) gives mass to up-type quarks. Thus, the model would have two \textbf{HCL}: one for $d$-type quarks and leptons, and another one for up-type quarks. 

We have argued that the  \textbf{HCL}'s predicted in Multi-Higgs models may provide interesting information about such models.
Furthermore, it could be interesting to calculate the angle subtended by these lines with respect to the SM one, as this observable 
may help to discriminate between the SM and those models. 

In the multi-Higgs models under consideration, the Higgs boson couplings with fermions ($y_f$) are functions of the masses  $\hat{m}_f$, where 
 $\hat{m}_f=m_f/\upsilon$. In the plane $\hat{m}_f - y_f$, we can define the vectors $\vec{w}_{fi}= (\hat{m}_{fi}, y^K_{fi})$ for models of type \textit{K}. Then we build the vectors $\vec{r}^K_f=\vec{w}_{fi}-\vec{w}_{fj}$, for two types of fermions $i$ and $j$, which may be from different families or different flavors within the same family.

 In order to determine the opening  angle between the \textbf{HCL} associated with SM and model \textit{K}, we define the  SM vector $\vec{r}^{\text{SM}}_f = \vec{w}^{\text{SM}}_{fi} - \vec{w}^{\text{SM}}_{fj}=  (\hat{m}_{fi} - \hat{m}_{fj}, y^{\text{SM}}_{fi}- y^{\text{SM}}_{fj})$, and $\vec{r}^{K}_f =  (\hat{m}_{fi}- \hat{m}_{fj}, y^K_{fi}- y^{K}_{fj})$, for model \textit{K}. Then, by evaluating their inner product, one finds the following expression for the opening angle between those lines:

\begin{equation}\label{angle}
\cos \Psi_{K} = \frac{ ( \Delta \hat{m}^f_{ij})^2  + \Delta {y}^K_{ij}  \Delta {y}^{\text{SM}}_{ij}  }
{ [ ( \Delta \hat{m}^f_{ij})^2 +  (\Delta {y}^K_{ij})^2]^{1/2}\cdot  [( \Delta \hat{m}^f_{ij})^2+ (  \Delta {y}^{\text{SM}}_{ij})^2]^{1/2} }
\end{equation}
where: 
\begin{eqnarray}
\Delta \hat{m}^f_{ij} &=& \hat{m}_{fi}-\hat{m}_{fj}\\ \nonumber
 \Delta {y}^K_{ij}&=& y^K_{fi}-y^K_{fj}\\ \nonumber
 \Delta {y}^{\text{SM}}_{ij}&=&y^{\text{SM}}_{fi}-y^{\text{SM}}_{fj} \nonumber
\end{eqnarray}

Thus, we are ready to calculate $\Psi_K$ for the 2HDM of different types. But first we need to check that the parameters of those 
models are in agreement with the most up-to-date Higgs measurements, for this purpose we shall use the recent reports 
on the Higgs boson properties from the ATLAS collaboration \cite{Aad:2019mbh} (while CMS results are from \cite{Sirunyan:2018koj}).  In addition, in order to confront our analysis with future expectations, we shall also consider the analysis for the HL-LHC \cite{Cepeda:2019klc}. Table \ref{kappas_LHC&HL-LHC} shows the corresponding results of the fits for Higgs boson coupling modifiers $\kappa_{X}$ (using 
the $\kappa$-formalism). 

\begin{table}
\caption{Fit results for Higgs boson coupling modifiers $\kappa_{X}$ reported
by ATLAS and CMS collaborations and the expected results at HL-LHC. \label{kappas_LHC&HL-LHC}}
\begin{centering}
\begin{tabular}{c c c c}
\hline 
$\kappa_{X}$ & ATLAS \cite{Aad:2019mbh} & CMS \cite{Sirunyan:2018koj} & HL-LHC \cite{Cepeda:2019klc}\tabularnewline
\hline 
\hline 
$\kappa_{t}$ & $1.03_{-0.11}^{+0.12}$ & $0.98\pm{0.14}$ & $1.04\pm0.025$\tabularnewline
\hline 
$\kappa_{b}$ & $1.00_{-0.22}^{+0.24}$ & $1.17_{-0.31}^{+0.27}$ & $0.94\pm0.028$\tabularnewline
\hline 
$\kappa_{\tau}$ & $1.04_{-0.16}^{+0.17}$ & $1.02\pm0.17$ & $1.0\pm0.17$\tabularnewline
\hline 
$\kappa_{Z}$ & $1.07_{-0.10}^{+0.11}$ & $1.00\pm0.11$ & $1.01\pm0.011$\tabularnewline
\hline 
$\kappa_{W}$ & $1.04\pm{0.10}$ & $-1.13_{-0.13}^{+0.16}$ & $1.01\pm0.011$\tabularnewline
\hline 
$\kappa_{\mu}$ & $<1.63$ & $0.80_{-0.80}^{+0.59}$ & $0.58\pm0.042$\tabularnewline
\hline 
\end{tabular}
\par\end{centering}

\end{table}

By considering the Higgs boson data, we can determine the {\bf{HCL}} associated with the best fit for the couplings of the Higgs boson, 
and then calculate the angle subtended with respect to  the SM prediction. We obtain the value 
$\Psi_{ex} =  (1.5 ^{+6.2} _{-2.9 }) \times 10^{-4} $. This result shows a value consistent with 
zero, but in order  to reach a more solid conclusion, we will have to wait for future data from LHC-HL.

\section{The opening angle $\Psi_K$ in the 2HDM-I, II \label{Sec3}}

We can use the master  formula \eqref{angle} in order to evaluate the angles predicted in the 2HDM of type I and II. Within the 2HDM-I, the Higgs-fermion couplings is given by: $y^I_f= \eta_f y^{\text{SM}}= \eta_f \hat{m}_f$, with $\eta_f= \cos\alpha / \sin \beta$, then the expression for the opening angle $\Psi_I$ simplifies as :
\begin{equation}
\cos \Psi_{I} = \frac{1 + \eta_f }
{ \sqrt{2} \left(1 + \eta^2_f \right)^{1/2} }.
\end{equation}

\begin{figure}[h!]
\centering
 \includegraphics[scale=0.21]{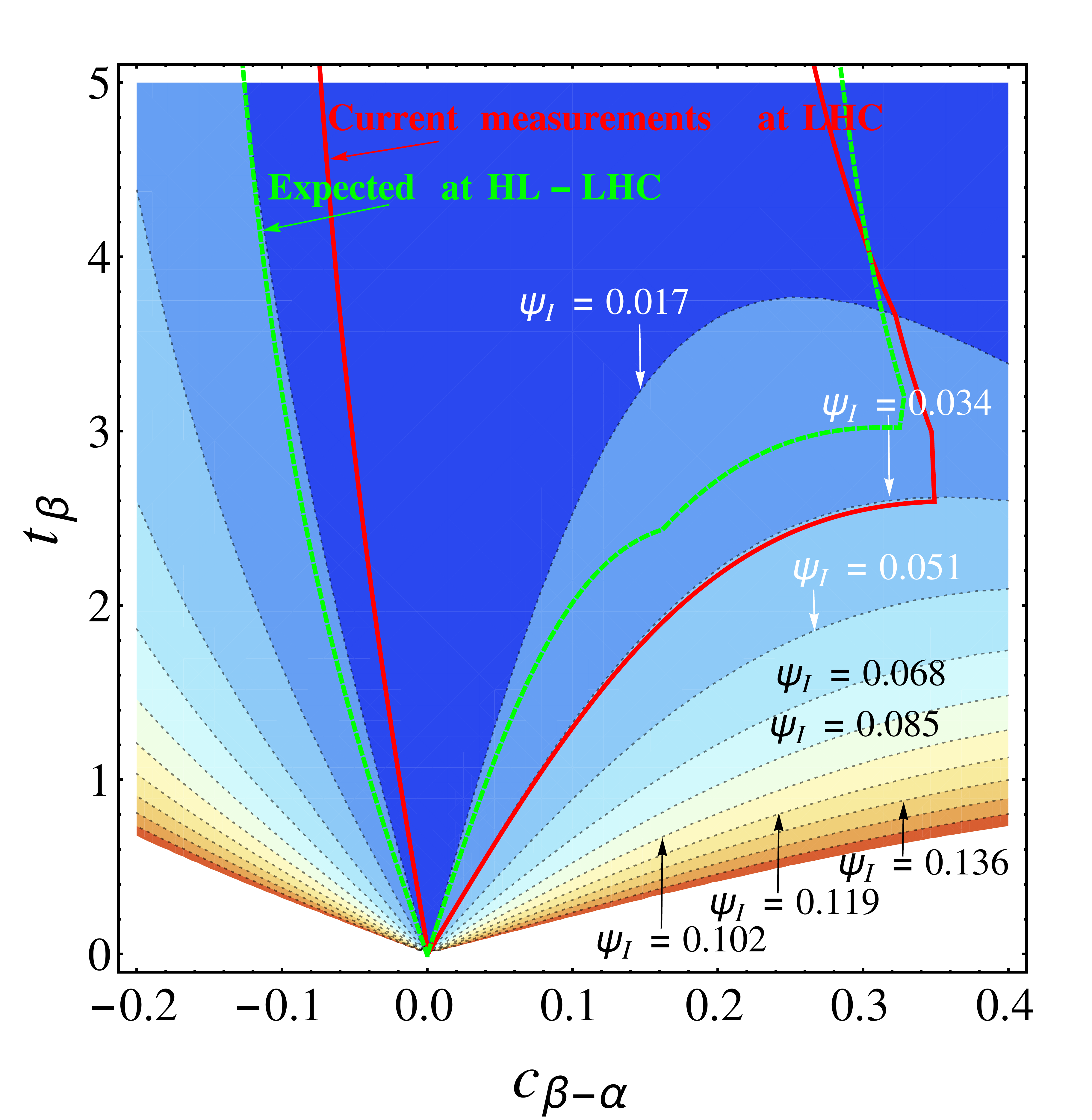}
\caption{
Values of the opening angle $\Psi_I$ with respect to the SM line for allowed regions in the plane $c_{\beta - \alpha}-t_{\beta}$ for 2HDM-I. }
\label{angles_2HDMI}
\end{figure}

 We show in Fig. \ref{angles_2HDMI}, the contour lines for the angle $\Psi_I$ within the 2HDM-I in the plane $c_{\beta-\alpha}-t_{\beta}$ (where 
  $\cos(\beta - \alpha)=c_{\beta-\alpha}$ and $\tan\beta=t_{\beta}$), as well as  the allowed  regions by LHC signal strengths $\mu_X$ with $X=\gamma,\,b,\,\tau,\,W,\,Z$. We assume couplings modifiers $\kappa_X$ with no-BSM contributions to the Higgs boson decays.  The allowed region by current LHC data corresponds to the area encircled by the red line; within that region we observe that $\Psi_I \lesssim 0.034$
  ($\simeq 2$ degrees). When the expected measurements at the HL-LHC are considered, the allowed region is now delimited by the dashed green line, in this case we find that $\Psi_I \lesssim 0.021$. 
  
\begin{figure}[h!]
\centering
\includegraphics[scale=0.24]{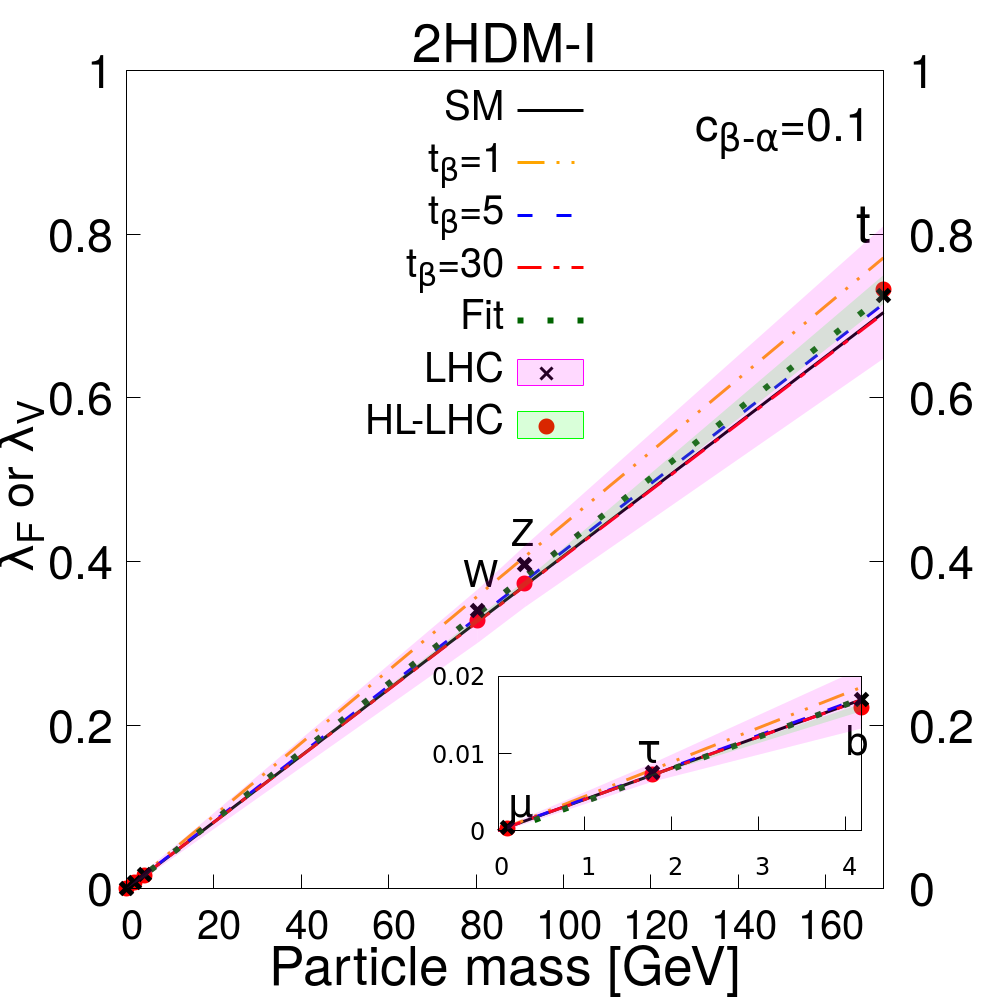}
\caption{
SM and 2HDM-I reduced coupling strength modifiers  $\lambda_{F}=\kappa_{F} \frac{m_F}{\upsilon}$ and $\lambda_{V}=\sqrt{\kappa_V}\frac{m_V}{\upsilon}$ with $F=t,\,b,\,\tau,\,\mu$ and $V=W,\,Z$ as a function of their masses. 
}
\label{yukawa-lines-I}
\end{figure}

  Then, we choose a few particular points in that plane, namely: $c_{\beta-\alpha}=0.1$ and $t_{\beta}=1, 5$ and $30$, and look for the plot of the   \textbf{HCL}. This is shown in Fig. \ref{yukawa-lines-I}, which includes the $\textbf{HCL}$ for both the SM and 2HDM-I cases. 
When the LHC results are considered (shaded pink area), it exclude values of $t_{\beta}\lesssim 1$, but values with $t_{\beta} > 1$ are allowed. The corresponding \textbf{HCL}'s  of the 2HDM-I  get closer to the SM line as $t_{\beta}$ grows. Figure \ref{yukawa-lines-I} also shows the best fit for the  $\textbf{HCL}$, which is separated from the SM one by about 2 degrees. A slight difference is observed when the bound on the signal strength $\mu_X$ is considered, in that case $t_{\beta}\lesssim 1.2$ is excluded. On the other hand, when the  HL-LHC option is considered we notice that a significant reduction of the possible allowed $\textbf{HCL}$ is obtained (shaded  green area).

For the 2HDM of type II, the $\textbf{HCL}$  associated with up-type quarks subtends an opening angle 
($\Psi^u_{II}$), while  $d$-type quarks and leptons define a second angle ($\Psi^d_{II}= \Psi^{\ell}_{II}$). 
The corresponding expressions for $\Psi^f_{II}$ are similar to eq. (4), but with $\eta_f$ replaced by  $\eta_u= \cos\alpha / \sin \beta$ and
$\eta_d= - \sin\alpha / \cos \beta= \eta_{\ell}$. 
We also notice that in the alignment limit, when $\tan \beta \simeq 1$, and $\alpha \simeq - \pi/4$, 
one gets: $\cos \Psi^f_{II} \simeq  1$, i.e. $\Psi^f_{II}=0$. 
 For up-type quarks it is not possible to get a value of the opening angle $\Psi^u_{II}$, as so far we only have
a precise measurement of the top quark Yukawa coupling in the up-quark sector.
However, for the $d$-type quarks and leptons, we can combine the corresponding measurements in order 
to calculate  $\Psi^d_{II}= \Psi^{\ell}_{II}$.

\begin{figure}[h!]
\centering
 \includegraphics[scale=0.22]{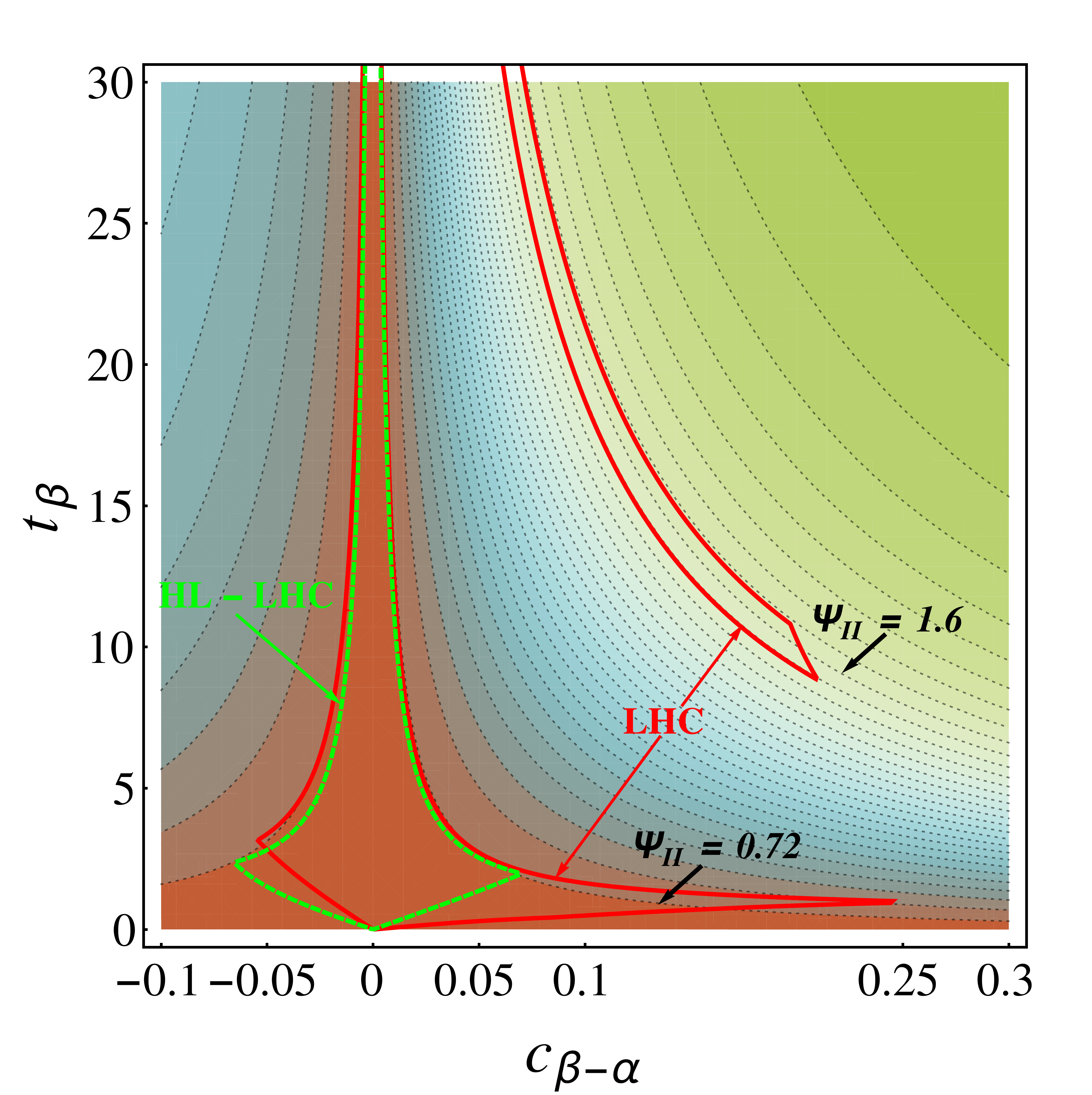}
\caption{
Values of the opening angle $\Psi_{II}$ with respect to the SM line for allowed regions in the plane $c_{\beta - \alpha}-t_{\beta}$ for 2HDM-II.}
\label{angles_2HDMII}
\end{figure}

\begin{figure}[h!]
\centering
\includegraphics[scale=0.23]{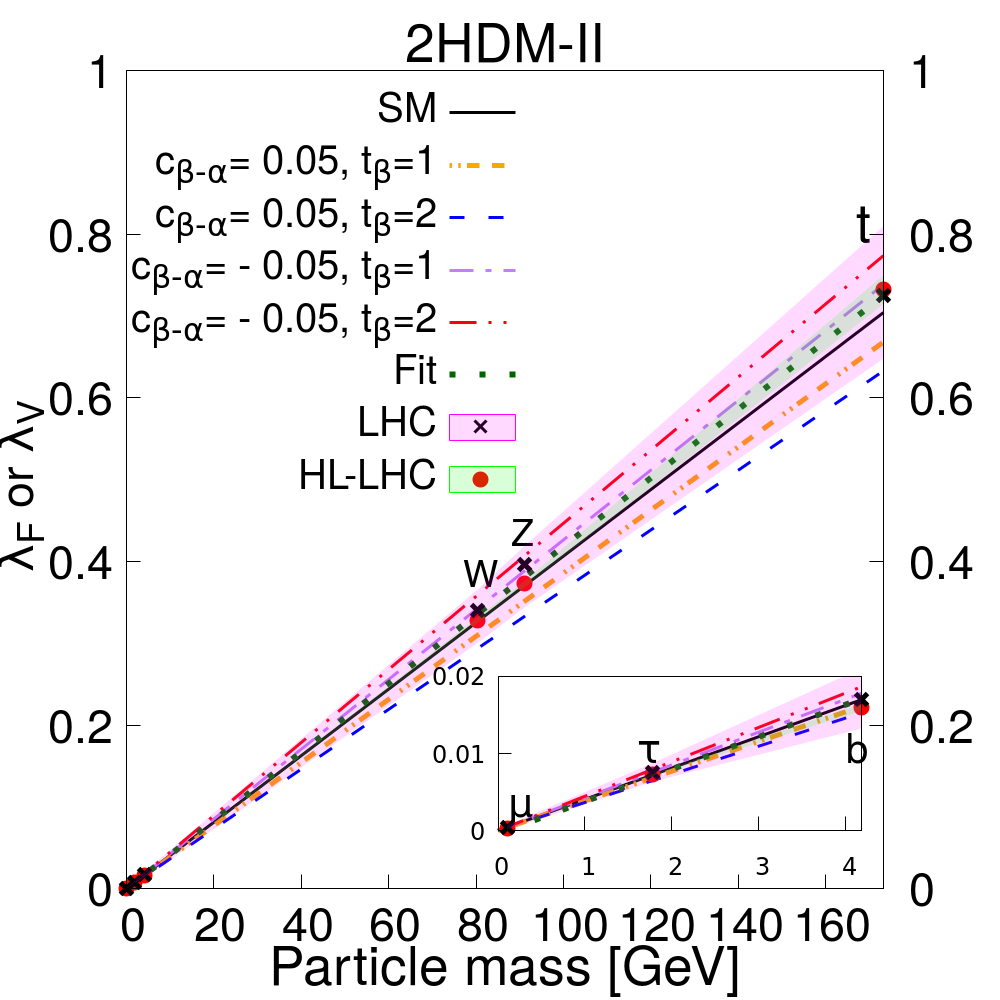}
\caption{
The same as the Fig. \ref{yukawa-lines-I} but the 2HDM-II.}
\label{yukawa-lines-II}
\end{figure}

The contour lines of the opening angle $\Psi^d_{II}$  are shown in Fig \ref{angles_2HDMII}, again in the plane $c_{\beta - \alpha}-t_{\beta}$, for points that satisfy all $\mu_{X}$ tests. Current LHC data allows two areas, which are encircled by the red lines, but one of them disappears
when the HL-LHC option is considered (dashed green lines). We also notice that when the region $c_{\beta-\alpha} \to 0$ is favored, it
 allows values up to $t_{\beta}=50$. Within the region close to $c_{\beta - \alpha} \simeq 0$, one find that $\Psi_{II} \lesssim 0.72$, whereas in the other area we have $\Psi_{II} \lesssim 1.6$. Then, in order to display the $\textbf{HCL}$,  we choose particular points in that plane, namely: 
\begin{itemize}
\item $c_{\beta - \alpha}=0.05$ and $\tan\beta= 1, 2$,
\item $c_{\beta - \alpha}=-0.05$ and $\tan\beta= 1, 2$. 
\end{itemize}
The  corresponding $\textbf{HCL}$'s for 2HDM-II are shown in Fig. \ref{yukawa-lines-II}.  
After considering  the LHC constraints (shaded pink area), we notice that all of the above points are allowed 
(small box with bottom, tau an muon couplings). 
 However, when one takes into account the HL-LHC option, the point $c_{\beta - \alpha}=0.05$ and $\tan\beta= 2$
 is no longer allowed, which shows how powerful are the constraints obtained from the opening angle
 in the 2HDM-II.

\newpage

\section{Higgs Couplings in effective theories of Flavor and neutrino masses\label{Sec4}}
It is also possible to construct models where some of the fermion masses are forbidden at the lowest order
 (dim-4 operators), and rather  involve higher-dimensional operators \cite{Babu:1999me}. In particular, one can use higher-powers 
 of the bilinear Higgs field, i.e. $(\Phi^{\dagger} \Phi)^n$, with $n$ being some integer adjusted to reproduce the fermion mass.
 For the complete 3-family case one needs a set of integers aimed to explain the hierarchy of the 
 fermion masses and mixing angles.
 For a single fermion type, the effective Lagrangian can be written, and expanded, as follows:
 
 \begin{eqnarray}
\cal{L} &=&  \lambda_f ( \frac{\Phi^{\dagger} \Phi }{ \Lambda^2} )^n  {\bar{F}}_{L} \Phi   f_{R} + ...+ h.c.\\ \nonumber
              &\simeq&   m_f \bar{f}_L f_R + (2n+1) \frac{m_f}{v}  \bar{f}_L f_R h  + h.c. 
 \end{eqnarray}
where $ \lambda_f$ is an $\mathcal{O}$(1) coefficient, $\Lambda$ is the scale at which these operators are induced
from a UV-completed theory. In the second step, we identified $m_f = \frac{v}{ \sqrt{2} }  [ \lambda_f ( \frac{v^2}{2\Lambda^2} )^n ] $. 
The second term implies that the $\textbf{HCL}$ is given by:
$g_{hff}= (2n+1) g^{\text{SM}}_{hff}$. Thus, we can see that for $n=0$ one reproduces the
SM relation for the Higgs couplings and the fermion mass, which is certainly correct for the top quark case. 
But for the lighter fermions, powers $n \geq 1$ are needed for $\Lambda \simeq 1$ TeV, 
which then result in $\mathcal{O}$(1) deviation from the SM case. 
This idea was implemented in ref. \cite{Babu:1999me} for the 3-family case, and the values of $n$ employed are already excluded 
by LHC data. 

More realistic models of this type, which are in agreement with LHC, have been constructed in ref. \cite{Bauer:2015kzy}, but this achieved by
considering two Higgs doublets. In general the expressions for the Yukawa Couplings of the Higgs boson will have the same form as before, i.e. with the factors $(2n+1)$, but we can suppress them with the vevs $\upsilon_2 < \upsilon_1$, or $\tan\beta$. For instance in the model of type II considered in ref. \cite{Bauer:2015kzy}, the $\kappa$-factor for the $d$-type quarks is given by:
\begin{equation}
\kappa_{d_i} = n_{d_i}  \frac{\cos \alpha}{\sin \beta} - (n_{d_i} +1) \frac{\sin \alpha}{\cos \beta}.
\end{equation}
The $b$-quark mass is reproduced with $n_{d_3}=1$, but in order to satisfy current LHC bounds on deviations from the SM Higgs couplings
with $b$-quark, one  requires special values of $\alpha$ and $\beta$. By considering that the Higgs coupling with massive vector bosons $W$ and $Z$ deviates just by a little from the SM case, i.e. $\cos(\beta - \alpha)=0.1$ and $\tan\beta \simeq 1$, we have that the positive solution for $\alpha$ gives a value $\kappa_b \simeq 2$, which is clearly excluded by the data shown in Table II. However, the negative solution for $\alpha$ implies a value of  $\kappa_b \simeq 1.07$, which is allowed by the Higgs LHC data. We leave a detailed numerical analysis of this model for
the future, including the possibility of having flavor violation \cite{Arroyo:2013tna}.



Thus, so far we have seen that the Higgs couplings with quarks and charged leptons display a  linear relationship 
with  the fermion masses. But, is it the only possibility?
What happens in the case of neutrinos? Here it depends on the type of  neutrino masses.  When their masses are of the Dirac type,
the corresponding relationships are also of the linear type. However, when neutrinos are of Majorana type, and its smallness is associated with
the see-saw mechanism, it is possible to have a non-linear relationship. We can illustrate this by considering a single family,
with a Lagrangian that includes both a Yukawa couplings of the type: $y_{\nu} \bar{L} \Phi \nu_R$, as well as a Majorana mass term for the right-handed neutrinos, $M \nu^c_R \nu_R$. Then the well known see-saw mass formula gives the light neutrino mass of the form:
$\frac{(y_\nu v)^2}{M}$, which then implies that the light neutrino coupling with the Higgs is given by:
$g_{h\nu\nu}= \frac{(m_\nu M)^{1/2}}{v}$. Thus, for neutrinos its coupling with the Higgs goes like the square root of the
neutrino mass, i.e. in this case we have a non-linear relation.




\section{Conclusions\label{Conclusions}}
Although the LHC has provided bounds on the new physics scale ($\Lambda$), 
that are already entering into the multi-TeV range, some of the motivations for new physics are  
so deep, that it seems reasonable
to wait for the next LHC runs, with higher energy and luminosity,
in order to have stronger limits, both in the search for new particles, such as heavier
Higgs bosons, and for precision tests of the SM properties.
The Higgs coupling with fermions and gauge bosons, as a function of the particle mass, 
are predicted  to lays on a single straight line within the SM. So far, the LHC has measured these couplings for the heavier
particles, which are found to satisfy this prediction,  but a complete test of this property must  also involve
 the lighter generations.  
 
 We have studied in this letter models with an extended Higgs sector, where there could be one or more Higgs Couplings lines ($\textbf{HCL}$), 
 such as the 2HDM or the Private Higgs Models.
We calculated the angle $\Psi_K$ subtended by those 
lines with respect to the SM one, using the values of the Higgs couplings obtained at LHC.
By considering the Higgs data from two flavors ($t$ and $b$, for instance) we can calculate
the angle subtended by the SM prediction, with respect to the best fit line, and we obtained the value 
$\Psi_{ex} =  (1.5 ^{+6.2} _{-2.9 }) \times 10^{-4} $, which is consistent with 
zero, but in order  to reach a more solid conclusion, we will have to wait for future data from LHC-HL.  
We also studied the predictions for these angles for several versions of the 2HDM, which are compared with the values of these angles
obtained using current data on the fermion masses and couplings. Our results indicate that LHC dat already constraints significantly the
parameters of the models. 
Models where the fermion masses hierarchy arise from higher-dimensional operators defined at the electroweak scale, 
may also result in large deviations from the SM values for the Higgs couplings, but in this case we would have an irregular pattern 
for the couplings, as a function of the fermion masses. In the case of neutrino masses, it happens that when the masses
 arise from the see-saw mechanism, the corresponding neutrino-Higgs couplings display a non-linear mass relation.
 Thus, by digging into the Higgs couplings it is possible to look for hidden patterns as the signals of Physics Beyond the SM. 

\begin{acknowledgments}
We would like to acknowledge the support of SNI (Mexico), and one of us (J.L.D.-C.) would like to thank VIEP(BUAP)
and the hospitality and support of the MCTP-UNACH for a sabbatical year. Discussions with Mario A. Perez de Leon are
also appreciated.
\end{acknowledgments}

\end{document}